\documentclass[%
 reprint,
superscriptaddress,
 amsmath,amssymb,
 prl,
]{revtex4-2}

\usepackage{graphicx}
\usepackage{dcolumn}
\usepackage{bm}

\usepackage{braket}
\usepackage{graphicx}
\usepackage{amsmath}
\usepackage{mathrsfs}
\usepackage{xcolor}
\usepackage{longtable}
\usepackage{amsfonts}
\usepackage{tikz}
\usepackage[colorlinks,linkcolor=blue,citecolor=blue,anchorcolor=blue]{hyperref}

\usepackage{CJK}
\usepackage{upgreek}
\usepackage{tikz}
\usetikzlibrary{quantikz}
\usepackage{xr}

\usepackage{amsmath}
\usepackage[normalem]{ulem}

\begin{document}

\preprint{APS/123-QED}

\title{Fault-Tolerant Connection of Error-Corrected Qubits with Noisy Links}

\author{Joshua Ramette}
\author{Josiah Sinclair}
\affiliation{%
 Department of Physics, MIT-Harvard Center for Ultracold Atoms and Research Laboratory of Electronics, Massachusetts Institute of Technology, Cambridge, Massachusetts 02139, USA
 }%
\author{Nikolas~P.~Breuckmann}
 \affiliation{%
 School of Mathematics, University of Bristol, Fry Building, Woodland Road, Bristol BS8 1UG, UK
 }%
 \author{Vladan Vuleti\'c}
\affiliation{%
 Department of Physics, MIT-Harvard Center for Ultracold Atoms and Research Laboratory of Electronics, Massachusetts Institute of Technology, Cambridge, Massachusetts 02139, USA
 }%

\date{\today}

\begin{abstract}
One of the most promising routes towards scalable quantum computing is a modular approach.
We show that distinct surface code patches can be connected in a fault-tolerant manner even in the presence of substantial noise along their connecting interface.
We quantify analytically and numerically the combined effect of errors across the interface and bulk.
We show that the system can tolerate 14 times higher noise at the interface compared to the bulk, with only a small effect on the code's threshold and sub-threshold behavior, reaching threshold with $\sim 1 \%$ bulk errors and $\sim 10 \%$ interface errors.
This implies that fault-tolerant scaling of error-corrected modular devices
is within reach using existing technology.
\end{abstract}

\maketitle

Quantum devices made from noisy components require error correction \cite{shor1995,Muralidharan2016} to scale.
Building error-corrected devices involves connecting a large number of qubits with gates of sufficiently high fidelity~\cite{Fowler2012}.
However, due to the general difficulty of controlling ever larger numbers of qubits within a single physical unit, quantum hardware platforms encounter practical system size limits.
For example, limits in the range of $10^2$-$10^4$ physical qubits are expected for trapped ions (due to spectral crowding of motional modes \cite{cetina2020}), superconducting qubits (due to cryostat size and chip fabrication \cite{Ang2022}), and Rydberg arrays (due to finite laser power and microscope field of view \cite{saffman_review_2016, Saffman2019}).
Because of these size limits, individual quantum processors may soon support multiple logical qubits \cite{ionq_web,ibm_web,ibm_web2,google_web,google_web2, Bluvstein2022}, but still not be truly scalable error-corrected devices.

To scale beyond these limits, one can consider architectures of local modules linked together via a physically distinct mechanism which is generally noisier and slower, e.g., trapped-ion chains connected via entangled photons \cite{Monroe2014}.
For a quantum computer, such a modular approach reduces the task of achieving true scalability to designing a unit module of fixed qubit number equipped with a fault-tolerant quantum input/output interface, so that scaling simply involves connecting more identical modules.
For quantum communication, a similar modular approach is already necessary to move quantum information between processors separated by long distances \cite{Muralidharan2016}.

\begin{figure}[]
\includegraphics[width=8.6cm]{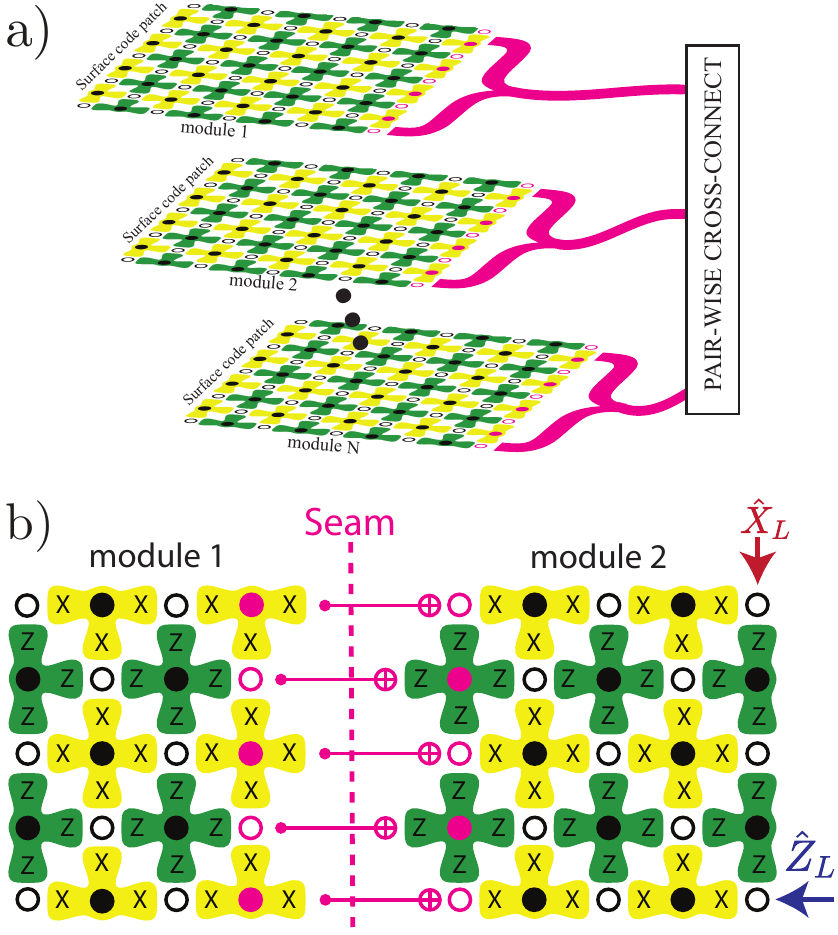}
\caption{a) Surface code qubits in distinct modules connected with noisy quantum communication links (pink) routed through a cross-connect switch.
b) A logical gate between two surface code ``bulk" patches in separate modules connected along a lower dimensional ``seam."
Stabilizer checks span the seam with pink teleported gates.
Pink data (open circle) and syndrome (filled circle) qubits lying along the interface of the two code patches experience elevated noise levels.
$\hat{X}_L, \hat{Z}_L$ indicate logical string operators.}
\label{pinkseam}
\end{figure}

A major challenge for error-corrected modular architectures is transferring quantum information between modules with sufficient speed and fidelity to satisfy the requirements for fault-tolerance. 
Because inter-module communication channels are typically lossy, quantum communication is accomplished by post-selected entanglement distribution. Entanglement shared between two modules in the form of nonlocal Bell pairs then serves as a resource to enable teleported gates
for inter-module operations \cite{Gottesman1999, Moehring2007, Stephenson2020}.

As even the heralded Bell pairs may still be noisy \cite{Stephenson2020, Nigmatullin2016}, entanglement distillation has been proposed to convert many low-fidelity pairs into a smaller number of higher-fidelity pairs \cite{Dur2003}.
However, simple distillation protocols can only reach errors approximately ten times larger than the local gate errors, because they involve many local gates \cite{Campbell2007, Krastanov2019}.
More sophisticated protocols are necessary to further reduce the noise, but the added complexity lowers the success rate,
reducing the achievable code cycle rate and increasing the memory errors per code cycle
\cite{Liang2007, Li2012, Fujii2012, Nickerson2013, Nickerson2014, Nigmatullin2016}.

\begin{figure}[]
\includegraphics[width=8.6cm]{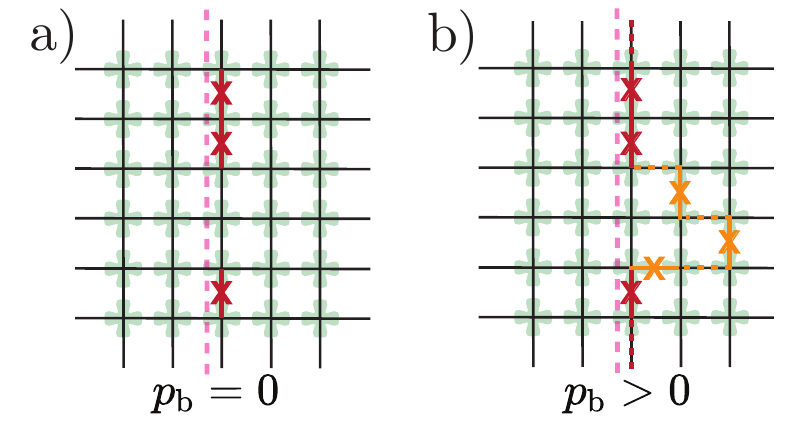}
\caption{Seam and bulk bit flip errors simultaneously contribute to logical failure. Edges are data qubits and vertices are check operators. a) Example of a correctable set of bit flips occurring only along the seam (vertical edges to the right of the pink line). b) With the same bit flips on the seam (red), bulk errors $p_\textrm{b}>0$ can form an ``excursion" (orange). Recovery now fails, filling in the dashed edges.}
\label{hops}
\end{figure}

In this Letter, we show that logical qubits encoded in surface code patches in distinct modules can be fault-tolerantly connected despite substantially elevated noise along their shared interface (see Fig.~\ref{pinkseam}a). 
While previous work pointed out that fault-tolerant quantum communication between code patches with noisy links is possible in the limit where the local noise in the bulk is asymptotically below threshold \cite{fowler2010}, we instead study the code in the presence of noise throughout both the bulk and interface, as in Fig.~\ref{hops}. Developing an understanding of how bulk and interface noise contributions combine to form logical failure modes, we provide analytical bounds and numerical simulations to show that the threshold for interface noise is as high as $\sim 10\%$, even with bulk noise close to the usual surface code threshold of $\sim 1 \% $. 
This relaxed threshold for communication errors implies that hardware platforms, that are already close to realizing local logical qubits \cite{ionq_web,ibm_web,ibm_web2,google_web,google_web2,Bluvstein2022}, only require noisy interconnects to immediately scale without distillation, better local gates, or other time or space overheads.

The surface code \cite{kitaev1997,bravyi1998quantum,freedman2001projective, Dennis2002} is a Calderbank-Shor-Steane (CSS) code with a high circuit-level decoding threshold of $p_\textrm{bulk}^* \approx$ 1\% \cite{Fowler2012}.
In Fig.~\ref{pinkseam} for an $L \times L$ surface code, the 4-body parity check operators $Z^{\otimes4}$ and  $X^{\otimes4}$, each realized by local CNOT gates between a syndrome qubit and four nearby data qubits, are indicated by the four green or yellow leaves associated with each syndrome qubit. The logical Pauli operators are strings of $X$ and $Z$ Pauli operators along the vertical and horizontal directions of the surface code, respectively.

Various methods exist to perform fault-tolerant two-qubit gates between surface code patches in separate modules. 
These include lattice surgery \cite{horsman2012surface}, braiding \cite{Fowler2012}, or even directly moving logical qubits between modules, but all reduce to maintaining a surface code patch spanning the two modules.
For logical computation across modules, it then suffices to connect the edges of two distinct surface code patches and perform parity checks spanning the modules which merge the two patches into a single larger surface code straddling the modules \cite{horsman2012surface}.
As shown in Fig.~\ref{pinkseam}b, along the interface or ``seam" where the two code patches connect, one leaf from each check stretches between the code patches, indicating a (pink) CNOT gate between a syndrome qubit and a data qubit in separate modules.

Making contact with methods to realize the inter-module gates (highlighted in pink in Fig.~\ref{pinkseam}b) via gate teleportation, Fig.~S1 in the Supplement \cite{supmat} shows how Bell pair bit and phase flip noise propagates onto the target and control qubits of the teleported gate, respectively.
Inspection of Fig.~\ref{pinkseam}b shows that  bit flip noise from the noisy Bell pair propagates to the right of the pink seam, and phase flip noise to the left of the pink seam.
As the $X$ and $Z$ errors trigger distinct check operators and because they can be decoded with distinct minimum-weight perfect matching (MWPM) decoders \cite{Dennis2002}, each decoder sees elevated noise on only a single strip of data+syndrome qubits to the right or left of the pink seam.

If the syndrome qubits were noiseless, the connecting interface would be equivalent to a $1$D repetition code (bit and phase flip noise rates $p_\textrm{s}$) embedded in the $2$D surface code (bit and phase flip noise rates $p_\textrm{b}$). 
To treat syndrome noise, we consider a generic phenomenological noise model where the bulk syndrome noise is set to be equal to the bulk data qubit noise $q_\textrm{b} = p_\textrm{b}$ \cite{Dennis2002}, and similarly for the seam syndrome and data qubit noise $q_\textrm{s} = p_\textrm{s}$, since Bell pair noise propagates to both.
Noisy syndrome decoding is accomplished through $L$ rounds of syndrome extraction, extending the matching graph along the syndrome qubits into an extra dimension representing time \cite{Dennis2002, Fowler2012}.
This results in a $D_\textrm{s} = (1+1)$ dimension lattice  of size $L \times L$ on which errors occur with rate $p_\textrm{s}$ embedded in a $D_\textrm{b} = (2+1)$ dimension lattice of size $L \times L \times L$ with error rate $p_\textrm{b}$.

We now show how the seam, due to its lower dimension than the bulk, can tolerate elevated levels of noise without severely compromising the integrity of the code spanning the modules, even in the presence of noise near the threshold $p_\textrm{b}^*$ within the bulk of the code.
Intuitively, the threshold of a surface code is determined by both the qubit noise level (the probability to extend a chain along a particular edge) and the number of directions available in which to extend the error chain.
In a lower dimension, fewer directions are available in which to extend error chains, resulting in a higher noise threshold.
Because $D_\textrm{b} > D_\textrm{s}$, the bulk and seam thresholds $p_\textrm{b}^*, p_\textrm{s}^*$ then satisfy $p_\textrm{s}^* > p_\textrm{b}^*$.

\begin{figure*}
\includegraphics[width=18cm]{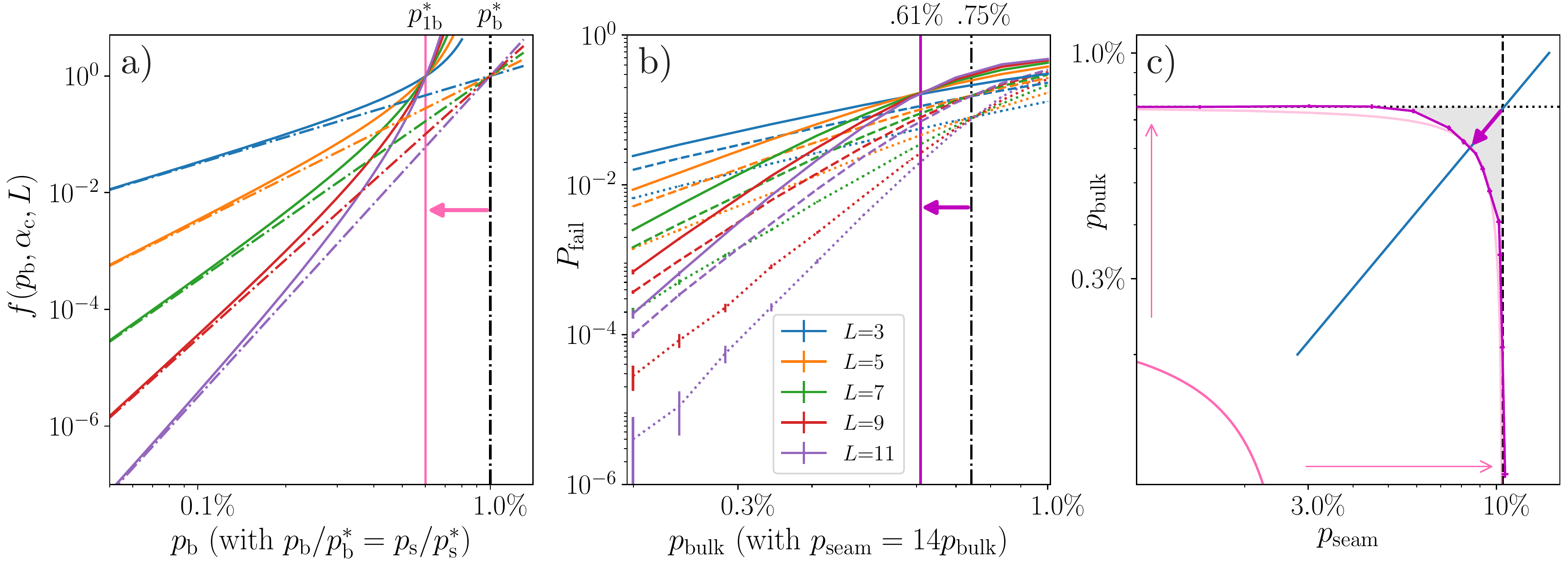}
\caption{
Analysis of the case $D_\textrm{b} = 3$ and $D_\textrm{s} = 2$.
a) Analytical bounds (Eq.~\ref{Pfail_equal}) for $p_\textrm{b}/p_\textrm{b}^* = p_\textrm{s}/p_\textrm{s}^*$ (solid), in which case the seam and bulk curves (dot-dashed) now overlap when plotted against $p_\textrm{b}$. Seam-bulk interactions reduce the threshold bound slightly to $p_\textrm{1b}^*$ as indicated by the pink arrow. The logical failure rate converges to the values for no seam-bulk interactions once the bulk error is a few times below $p_\textrm{b}^*$, as excursions into the bulk become ``frozen out."
b) Same as a) but exact numerical simulation with the choice $p_\textrm{seam} = 14 p_\textrm{bulk}$, approximately $p_\textrm{bulk}/p_\textrm{bulk}^* = p_\textrm{seam}/p_\textrm{seam}^*$, which aligns the thresholds for the bulk-only curves (dotted) and seam-only curves (dashed). The full simulation including seam-bulk interaction (solid) similarly sags slightly and then converges toward the seam-only curves as $p_\textrm{bulk}/p_\textrm{bulk}^*$ becomes small.
c) Numerically extracted threshold plotted in terms of $p_\textrm{seam}, p_\textrm{bulk}$ (purple).
Pink shows the threshold bound (Eq.~\ref{thresh_formula}), with arrows pointing to the light pink line showing how the bound formula relaxes when reducing the overcounting of paths by substituting numerical values for $p_\textrm{seam}^*$ and $p_\textrm{bulk}^*$ along with an effective value of $\alpha_\textrm{c} \rightarrow 1.4$, the minimal value still bounding all the numerical datapoints.
}
\label{threshold_shifts}
\end{figure*}

Next, we consider how the extra seam noise in Fig.~\ref{pinkseam}b affects the probabilities for $\hat{X}_L$ and $\hat{Z}_L$ logical errors.       
Since logical $\hat{X}_L$ errors can occur both in the bulk and along the seam, suppressing them requires both seam noise below seam threshold, $p_\textrm{s} < p_\textrm{s}^*$, and bulk noise below bulk threshold, $p_\textrm{b} < p_\textrm{b}^*$.
This is the most stringent case and considered in Fig.~\ref{threshold_shifts} and the remainder of the paper.
In contrast, even for $p_\textrm{s} > p_\textrm{s}^*$, a $\hat{Z}_L$ error still must penetrate the length of the bulk and can be suppressed so long as the bulk is below threshold.
For equal data and syndrome qubit noise as we consider ($q_\textrm{b} = p_\textrm{b}$ and $q_\textrm{s} = p_\textrm{s}$), the probability of generating ``time-like" error chains stretching through the $L$ rounds of error correction for both the bit and phase flip decoders is identical to that for $\hat{X}_L$.
Thus, Fig.~\ref{threshold_shifts} also shows the probability for time-like bit and phase flip error chains spanning $L$ rounds. 
However, once the threshold criteria from Fig.~\ref{threshold_shifts} is reached, these are easily suppressed by extending the code further in the time direction with more rounds of error correction per logical gate.

Now, in order for a logical bit-flip error to occur, the combined effect of a round of errors and corrections must generate some nontrivial chain of bit flips $\{ \gamma \}$ of length at least $L$.
First considering a single code patch with no seam, as shown in \cite{breuckmann2018, Dennis2002, dumer2015, strikis2021} and outlined in the Supplement \cite{supmat}, the logical failure probability $P_\textrm{fail}$ is bounded by the number of such possible chains $n_\textrm{nontrivial}(L)$ times the probability for each to occur:
\begin{equation}
    P_\textrm{fail}/\textrm{poly}(L) \leq n_\textrm{nontrivial}(L) \times (2^L p^{L/2})
    \label{Pfail_homogenous}
\end{equation}

Here, the probability bound of $2^L p^{L/2}$ originates from how MWPM fills in the rest of a nontrivial chain once $L/2$ bits flip due to environmental noise, and there being $\leq 2^L$ ways to choose half or more of the $L$ bits to have flipped.
To express a bound on $n_\textrm{nontrivial}(L)$, consider how when appending each additional edge to $\{ \gamma \}$, one can move in any direction on the lattice other than back, so in $2D-1$ directions in dimension $D$, bounding $n_\textrm{nontrivial}(l) \leq (2D-1)^l \equiv \mu_D^l$. We can then rewrite:
\begin{equation}
    P_\textrm{fail}/\textrm{poly}(L) \leq (4 \mu_D^2 \times p)^{L/2} \equiv \Big(\frac{p}{p^* } \Big)^{L/2}
    \label{bound_homogenous_final}
\end{equation}
which is exponentially suppressed to zero as $L \rightarrow \infty$ for $p < p^*$. As $\mu_D$ is dimension dependent, so is the threshold bound $p^* \equiv 1/(4 \mu_D^2)$.

In the Supplement \cite{supmat}, we derive bounds (Eqs. \ref{seam+excursions} - \ref{Pfail_equal}) constraining how the threshold changes in the presence of a noisy seam by counting the number of error chains which hop between the seam and bulk. 
Here, we present a heuristic argument reaching the same conclusions and elucidating the failure mechanisms.
We can interpret our expression for the logical failure (Eq.~\ref{bound_homogenous_final}) as the factor of $\mu_D \times 2 \sqrt{p}$ for each additional edge appended to $\{ \gamma \}$ (so a probability of $(\mu_D \times 2 \sqrt{p})^L$ for appending $L$ edges), with $\mu_D$ ways to add edges, and $2 \sqrt{p}$ as an effective ``probability" per link (with the factor of 2 and the square root appearing because of how the MWPM procedure fills in missing links).
If errors occur solely along the seam with $p_\textrm{b} = 0$ as in Fig.~\ref{hops}a, we would have a factor of $\mu_\textrm{s} \times 2 \sqrt{p_\textrm{s}}$ per appended edge.
But making $p_\textrm{b} > 0$, we suddenly allow additional paths through the bulk before appending the next seam link as in Fig.~\ref{hops}b, which adds terms corresponding to excursions into the bulk before reattaching to the seam.
For a given excursion, in addition to flipping an arbitrary possible number of bulk edges $\ell$, each of which can be appendded in $\mu_\textrm{b}$ ways for a factor of $( \mu_\textrm{b} \times 2\sqrt{p_\textrm{b}} )^{\ell}$, an excursion also involves flipping two bulk edges orthogonal to the direction along the seam as well as the final seam link for another factor of $\mu_\textrm{c} \times 2 \sqrt{p_\textrm{s}} (2 \sqrt{p_\textrm{b}})^2$.
The factor $\mu_\textrm{c} \equiv 4 D_\textrm{s} (D_\textrm{b}-D_\textrm{s})$ counts the number of ways to choose both how to leave $2(D_\textrm{s} - D_\textrm{b})$ and reattach $2 D_\textrm{s}$ to the seam.
By analogy with Eq.~\ref{bound_homogenous_final}, we add up all of these ways to attach new seam links, giving us a modified factor for each seam edge we append:
\begin{align}
    \mu_\textrm{s} \times 2 \sqrt{p_\textrm{s}}  + &  \sum_{\ell=0}^\infty \Big(\mu_\textrm{c} \times 2 \sqrt{p_\textrm{s}} (2 \sqrt{p_\textrm{b}})^2 \Big) \Big( \mu_\textrm{b} \times 2\sqrt{p_\textrm{b}} \Big)^{\ell} \nonumber \\
    & = \sqrt{ \frac{p_\textrm{s}}{p_\textrm{s}^*}} \Bigg(1 + \alpha_\textrm{c} p_\textrm{b} \frac{\sqrt{p_\textrm{s}^*}}{1-\sqrt{p_\textrm{b}/p_\textrm{b}^*}}\Bigg)
    \label{seam+excursions}
\end{align}
where we have summed the geometric series over $\ell$ and defined $\alpha_\textrm{c} \equiv 8 \mu_\textrm{c}$.

Again by analogy, we would then expect the failure probability bound to be Eq.~\ref{seam+excursions} raised to the power of $\gamma_S$, the number of edges in $\{ \gamma \}$ on the seam.
Additionally, realizing  that if $\gamma_S < L$ there must be at least $L - \gamma_S$ bulk links in $\{ \gamma \}$, as $\{ \gamma \}$ must have at least $L$ total links in order to fail, we arrive at:
\begin{multline}
    P_\textrm{fail}/\textrm{poly}(L) \leq
    \Big(\frac{p_\textrm{s}}{p_\textrm{s}^*} \Big)^{\frac{L}{2}} + \Big(\frac{p_\textrm{b}}{p_\textrm{b}^*}\Big)^{\frac{L}{2}} + \\
    \sum_{\substack{ \gamma_S\geq 1: \\ C \neq 0}}^{L} \Bigg[ \frac{p_\textrm{s}}{p_\textrm{s}^*}\Big(1 + \alpha_\textrm{c} p_\textrm{b} \frac{\sqrt{p_\textrm{s}^*}}{1-\sqrt{p_\textrm{b}/p_\textrm{b}^*}}\Big)^2 \Bigg]^{\frac{\gamma_S}{2}} \Bigg[\frac{p_\textrm{b}}{p_\textrm{b}^*}\Bigg]^{\frac{L-\gamma_S}{2}}
    \label{Pfail_cross}
\end{multline}

The first two terms correspond to failure chains within purely the seam or bulk and the additional cross terms apply when considering chains with excursions ($C \neq 0$), in which case the threshold criteria mixes conditions on $\frac{p_\textrm{s}}{p_\textrm{s}^*}$ and $\frac{p_\textrm{b}}{p_\textrm{b}^*}$.
All these terms are suppressed as $L \rightarrow \infty$ provided that the quantities in brackets are smaller than unity.
Re-expressing Eq.~\ref{seam+excursions},
we can see that it is equivalent to a small downward ``sag" of the seam threshold bound:
\begin{equation}
    p_\textrm{s}^* \rightarrow p_\textrm{1s}^* \equiv p_\textrm{s}^* \bigg/ \Bigg[1 + \alpha_\textrm{c} p_\textrm{b} \frac{\sqrt{p_\textrm{s}^*}}{1-\sqrt{p_\textrm{b}/p_\textrm{b}^*}} \Bigg]^{2}
    \label{thresh_formula}
\end{equation}

Fixing $\frac{p_\textrm{s}}{p_\textrm{s}^*} = \frac{p_\textrm{b}}{p_\textrm{b}^*}$, Eq.~\ref{Pfail_cross} reduces to:
\begin{multline}
    P_\textrm{fail}/\textrm{poly}(L) \leq \\
    \Bigg[ \frac{p_\textrm{b}}{p_\textrm{b}^*}\Big(1 + \alpha_\textrm{c} p_\textrm{b} \frac{\sqrt{p_\textrm{s}^*}}{1-\sqrt{p_\textrm{b}/p_\textrm{b}^*}}\Big)^2 \Bigg]^{\frac{L}{2}} \equiv f(p_\textrm{b}, \alpha_\textrm{c}, L)
    \label{Pfail_equal}
\end{multline}
which we plot in Fig.~\ref{threshold_shifts}a.

In Fig.~\ref{threshold_shifts}, we compare these analytical bounds to numerical Monte Carlo simulations \cite{qecsim, Higgott2021} to quantify the effect of error chains stretching simultaneously across the bulk and the seam.
An additional consequence of the seam being a sublattice within the bulk is that per code cycle, while bulk qubits are addressed by four local gates, each seam qubit only interacts with a single Bell pair for communication.
Letting $p_\textrm{bulk}$ represent the probability of a local gate in the bulk to cause a bit flip on a bulk qubit, we model this by directly substituting $4p_\textrm{bulk} = p_\textrm{b}$ (leading to a reasonable $p_\textrm{bulk}^* = 0.75\%$) while maintaining $p_\textrm{seam} = p_\textrm{s}$, since seam qubit noise is dominated by the Bell pair.
From Fig.~\ref{threshold_shifts}, we see that the numerical results display the same qualitative behavior as the analytical formulas, with a slight sag in the threshold as well as subthreshold convergence of the logical failure to the same value as without excursions. By adjusting the parameters $\mu_\textrm{s}$ and $\mu_\textrm{b}$ to match the known surface code threshold values for $D_\textrm{s} = 2$ and $D_\textrm{b} = 3$ \cite{Dennis2002} and relaxing $\alpha_c$ from the rigorous bound to a smaller effective value, the bound formulas also provide a fairly accurate analytical model of the threshold behavior.
Notably, up to the small correction represented by the gray region of Fig.~\ref{threshold_shifts}c, $P_\textrm{fail}$ behaves approximately as if bulk and seam were decoupled without cross terms from Eq.~\ref{Pfail_cross}:
\begin{equation}
    P_\textrm{fail}(p_\textrm{bulk},p_\textrm{seam}) \approx \Big( \frac{p_\textrm{bulk}}{p_\textrm{bulk}^*} \Big)^{L/2} + \Big( \frac{p_\textrm{seam}}{p_\textrm{seam}^*} \Big)^{L/2}
    \label{wrongmodel}
\end{equation}
with $p_\textrm{bulk}^* \approx 1 \%$ and $p_\textrm{seam}^* \approx 10 \%$.
As long as $\frac{p_\textrm{seam}}{p_\textrm{seam}^*} \approx \frac{p_\textrm{bulk}}{p_\textrm{bulk}^*}$, so $p_\textrm{seam} \approx 10 p_\textrm{bulk}$, the seam noise has almost no effect.

\begin{figure}
\includegraphics[width=9cm]{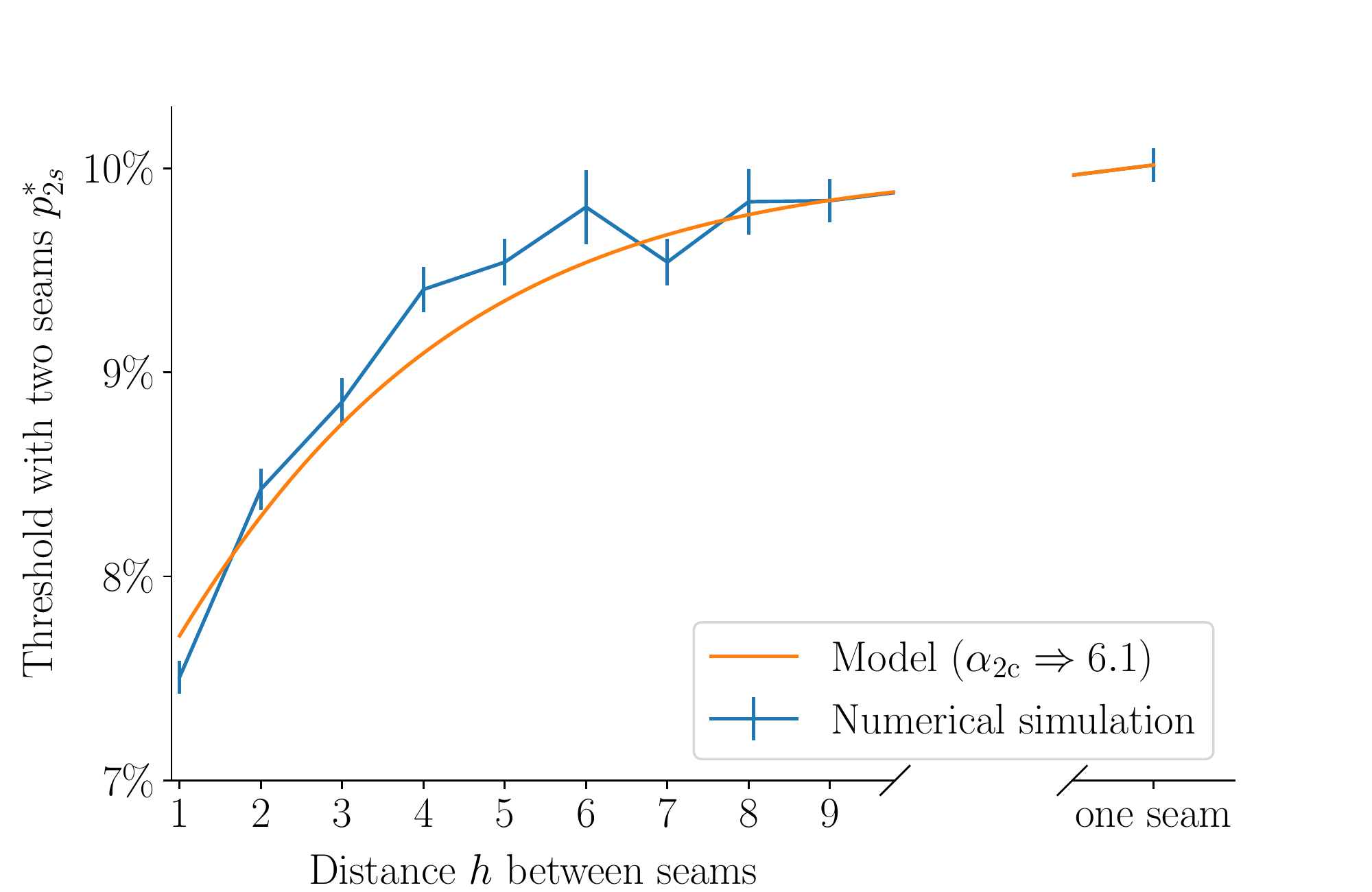}
\caption{
Effect on the threshold due to two nearby parallel seams, fixing $p_{\textrm{b}}/p_{\textrm{b}}^{*}= \frac{1}{2}$. Blue: numerical simulations. Yellow: Eq.~\ref{2_seam_sag} plotted with fit value $\alpha_\textrm{2c} = 6.1$. The closer the seams (smaller $h$), the easier it is to hop between seams, lowering the threshold. For large $h$, long excursions into the bulk are exponentially suppressed and the threshold returns to the value for just one seam.}
\label{2_seam}
\end{figure}

We further explain how our formalism can generalize to connecting surface code patches in a variety of configurations, such as transversal gates or a grid of smaller patches.
While in Fig.~\ref{pinkseam}b we aligned the physical edges of two code patches so that the seam extended in both space and time, the results from Fig.~\ref{threshold_shifts} can apply to any $2D$ subspace of a $3D$ lattice.
Consequently, a transversal gate between logical qubits in separate modules mediated by Bell pairs and followed by rounds of local error correction on each logical qubit would be similarly robust to Bell pair noise, as the transversal gate occurs on a single code cycle and introduces noise on a $2D$ sublattice which now extends in two space directions rather than one space and one time direction as before.

By adding multiple seams along space and/or time directions and counting the paths to hop between different seams, our formalism can be used to understand situations including repeated transversal gates and code patches spanning multiple modules.
As an illustration, we model two parallel $D_\textrm{s} = 2$ noisy seams embedded within the same $D_\textrm{b} = 3$ bulk, capturing the additional process of hopping between the two seams by adding to Eq.~\ref{seam+excursions} the term:
\begin{equation}
    \mu_\textrm{2c} \sqrt{4 p_\textrm{s}} \sqrt{4 p_\textrm{b}}^2 \sum_{\ell=h-2}^\infty \Big( \mu_\textrm{b} \sqrt{4 p_\textrm{b}} \Big)^{\ell}
\end{equation}
with chains of $\ell$ bulk edges hopping between seams. $\mu_\textrm{2c}$ represents an effective factor counting the number of ways to leave and reattach to distinct seams and $h$ is the distance separating the two embedded seams (the sum begins at $h-2$ because in addition to its two corner edges, the hop must traverse through $h-2$ bulk edges).
For a fixed ratio $p_\textrm{b}/p_\textrm{b}^*$, we obtain an expression in terms of $p_\textrm{1s}^*$:
\begin{multline}
    \sqrt{\frac{p_\textrm{s}}{p_\textrm{1s}^{*}}} + \mu_\textrm{2c} \sqrt{4 p_\textrm{s}} \sqrt{4 p_\textrm{b}}^2 \sum_{\ell=h-2}^\infty \Big( \mu_\textrm{b} \sqrt{4 p_\textrm{b}} \Big)^{\ell} \nonumber \\
    = \sqrt{\frac{p_\textrm{s}}{p_\textrm{1s}^{*}}} \Bigg[1 + \alpha_\textrm{2c} \sqrt{ p_\textrm{1s}^{*}} \sqrt{p_\textrm{b}}^2 \Big(\frac{p_\textrm{b}}{p_\textrm{b}^*} \Big)^{(h-2)/2} \frac{1}{1-\sqrt{p_\textrm{b}/p_\textrm{b}^*}} \Bigg] \nonumber
\end{multline}
with $\alpha_\textrm{2c} \equiv 8 \mu_\textrm{2c}$ From this we observe that the seam threshold further sags to $p_\textrm{2s}^{*}$ if the two seams are nearby:
\begin{align}
    p_\textrm{1s}^{*} \rightarrow p_\textrm{2s}^{*} \equiv
    p_\textrm{1s}^{*} \bigg/ \Bigg[1 + \alpha_\textrm{2c} \frac{\sqrt{p_\textrm{1s}^{*}} p_\textrm{b}}{1-\sqrt{p_\textrm{b}/p_\textrm{b}^*}} \Big(\frac{p_\textrm{b}}{p_\textrm{b}^*} \Big)^{(h-2)/2} \Bigg]^2
    \label{2_seam_sag}
\end{align}
which we plot in Fig.~\ref{2_seam}, revealing that larger $h$ suppresses hops between seams, leading to a smaller sag of the threshold.
The simulations in Fig.~\ref{2_seam} show a similar behavior and a quantitative match for an effective value of $\alpha_\textrm{2c} \rightarrow 6.1$.

Our main result, that the interface can tolerate a noise level $p_\textrm{seam} \approx 10 p_\textrm{bulk}$ (Eq.~\ref{wrongmodel}),
relaxes the communication fidelity required for fault-tolerance in ongoing experimental efforts to build modular architectures.
Such efforts include optical interconnects between ions \cite{Stephenson2020}, atoms \cite{Reiserer_2015}, or superconducting qubits \cite{Sahu2022, Tu2022, Delaney2022, Zhu_2020, Imany_2022}, direct superconducting microwave links \cite{Magnard2020, Burkhart2021, Yan2022, Zhong2021}, or even shuttling ions \cite{Pino2021} or atoms \cite{Bluvstein2022} between distinct subunits, some having already demonstrated communication errors below $p_\textrm{seam}^* \approx 10 \% $ \cite{Stephenson2020, Burkhart2021, Zhong2021}.
With multiple quantum computing platforms \cite{ionq_web,ibm_web,ibm_web2,google_web,google_web2, Bluvstein2022} rapidly progressing toward module sizes of thousands of qubits and local gate noise targets of $\sim 0.1\%$, interconnects with a corresponding target of $\sim 1 \%$ noise will directly enable fault-tolerant modular scalability.
For even noisier networks, distillation can still be used, with the simplest and most efficient protocols being sufficient to reach errors of $\sim 10 p_\textrm{bulk}$ \cite{Krastanov2019}, enabling scalability.

\section*{Acknowledgement}
This project was funded in part by DARPA under the ONISQ program (grant \# 134371-5113608), the MIT-Harvard Center for Ultracold Atoms (NSF grant \# PHY-1734011), DoE under the Quantum Systems Accelerator Center (contract \# 7571809), and AWS.

The authors thank Dolev Bluvstein for helpful discussions.

J.R. originally conceived the work, developed the theoretical formalism, and derived the bound results. J.R. and J.S. conceptually developed the work, performed and interpreted the numerical simulations, and drafted the manuscript. N.B. contributed to formalizing the bound proof and provided helpful discussions. V.V. supervised the project. All authors discussed the results and contributed to the writing of the manuscript.

\bibliography{ramette_biblio}

\end{document}


\preprint{APS/123-QED}

\title{Supplementary Material:\\
Fault-Tolerant Connection of Error-Corrected Qubits with Noisy Links}

\author{Joshua Ramette}
\author{Josiah Sinclair}
\author{Nikolas P.~Breuckmann}
\author{Vladan Vuleti\'c}

\date{\today}

\maketitle

\section{Network noise propagation for teleported gates}

\begin{figure}[h]
\includegraphics[width=0.5\textwidth]{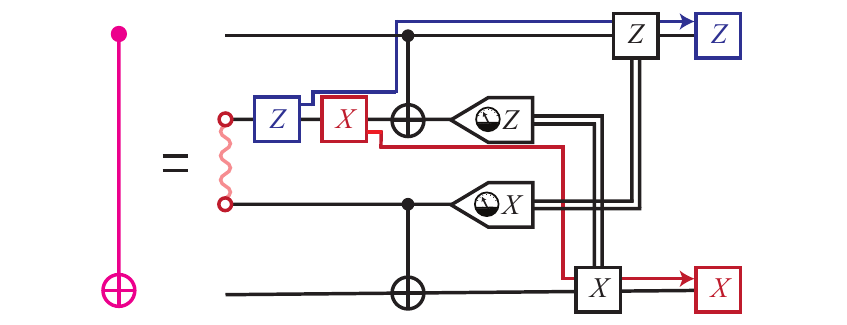}
\caption{$X$, $Z$ noise on a Bell pair (red squiggle) used in a teleported gate propagates to the two qubits it operates on. Phase flips only propagate to the control, and bit flips to the target.}
\label{circuit}
\end{figure}

When the interface is realized via distributed entanglement, that entanglement serves as a resource for enacting non-local, teleported gates between qubits in distinct modules.
Fig.~\ref{circuit} shows how bit and phase flip noise on the distributed Bell pair propagates to the control and target qubits in the distinct modules that the (pink) teleported gate acts on.
The propagation is identical for errors occurring on either of the Bell pair qubits, as must be the case since the Bell pair is invariant under application of $XX$ and $ZZ$.

\section{Bounds}
One can lower bound the phenomenological thresholds for surface codes by counting walks corresponding to homologically nontrivial error chains. A surface code with noiseless syndromes can be decoded by pairing up defects on a lattice in $2D$, and to include noisy syndromes this is simply extended to the problem of pairing up defects which additionally propagate in time as a $3D$ matching problem. Previous bounds in $2D$ and $3D$ \cite{Dennis2002}, while not tight, were within about a factor of 3 of the true thresholds.

To illustrate the main concepts of the paper concerning error chains in a ``bulk" matching graph lattice (dimension $D_\textrm{b}$) containing a ``seam" subspace lattice (dimension $D_\textrm{s} < D_\textrm{b}$), we count the numbers of walks and their probabilities, including those which span across both the bulk and the seam. For the cases of interest ($D_\textrm{b}$ = 2, $D_\textrm{s} = 1$ for noiseless syndromes and $D_\textrm{b}$ = 2+1, $D_\textrm{s} = 1+1$ for noisy syndromes), pairing defects on the matching graph allows decoding of the surface code of corresponding dimension.
When the bulk is operating slightly below its own threshold, we can quantify how the probability of long ``excursions" away from the seam is strongly suppressed.
We evaluate our results for the case of greatest interest ($D_\textrm{b} = 3$, $D_\textrm{s}=2$), showing it is possible to increase noise on the seam in exchange for a modest decrease of noise in the bulk.
While these bounds are not tight, and the combinatorics tracking the hopping between seam and bulk over-counts the walks to give less impressive results than revealed by the exact numerical simulations, the derivation illustrates many core concepts important to understand our results and also insight into more complex situations, such as how the code performance is affected by having multiple different seams operating within the same bulk.

\subsection{The matching graph and the MWPM decoder}
As shown in Fig.~2 of the main text, the set $\{ M \}$ of edges in the matching lattice corresponds to data qubits, and its vertices correspond to syndromes, such that the vertices forming the boundary $\partial$ of a set of errors on $\{ M \}$ correspond exactly to the violated checks.

\begin{table}[h]
    \centering
\begin{tabular}{| c | l |}
 \hline
 $\{ M \}$ & set of all edges in the matching graph  \\ 
  \hline
 $\{ E \}$  & set of all edges bit flipped due to noise\\
  \hline
 $\{ R \}$  & set of all edges flipped due to MWPM recovery \\
  \hline
$\{ S \}$  & set of all edges on the seam \\
  \hline
$\{ B \}$  & set of all edges in the bulk \\
  \hline
 $\{ \gamma \}$ & Set of connected edges on a closed path \\
   \hline
 $\{ \gamma_{S} \}$ & $\{ \gamma \} \cap \{ S \}$, set of edges in $\{ \gamma \}$ and on the seam \\
  \hline
 $\{ \gamma_{B} \}$ & $\{ \gamma \} \cap \{ B \}$, set of edges in $\{ \gamma \}$ and in the bulk \\
  \hline
 $\{ \gamma_{SE} \}$ & $\{ \gamma_S \} \cap \{ E \}$, set of seam noise bit flips in $\{ \gamma \}$ \\
  \hline
 $\{ \gamma_{SR} \}$ & $\{ \gamma_S \} \cap \{ R \}$, set of seam recovery bit flips in $\{ \gamma \}$ \\
  \hline
\end{tabular}
    \caption{A handy glossary of terms.}
\end{table}

\begin{figure}[h]
\includegraphics[width=8.6cm]{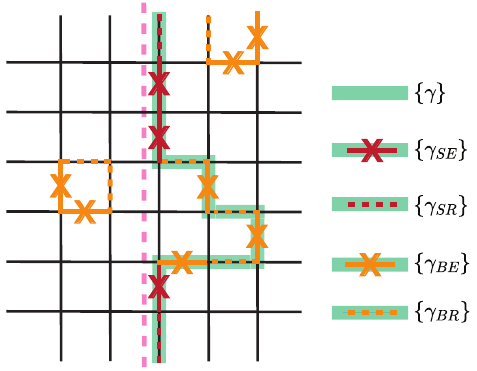}
\caption{
Logical failure occurs if during a round of error correction, enough bits are flipped by environmental noise combined with the attempted correction to form some nontrivial chain $\{ \gamma \}$ spanning the code.
Each round, errors introduce a random set $\{ E \}$ of bit flips, which occur both on seam edges and on bulk edges as red and orange X's, respectively.
MWPM recovery further bit flips the dashed edges $\{ R \}$ on the seam and the bulk to return the state to the codespace.
Then, the remaining X's and dashes together for the set $\{E + R \}$.
In this example, $\{ E + R \}$ contains the nontrivial chain $\{ \gamma \}$.
}
\label{pinkseam}
\end{figure}

We first establish some terminology. ``Edges" refers to locations of qubits in the matching graph.
Quantities in brackets $\{ \}$ refer to matching graph subsets, and quantities without brackets refer to the sizes of such subsets.
$\{E\}$ is a (possibly disconnected) set of edges where bit flip errors occurred on a given round of error correction, and $\{R\}$ is a set of edges chosen via minimum weight perfect matching (MWPM) which are bit flipped to attempt correction of $\{E\}$.
The sum of the error and recovery steps $\{E + R\} \equiv (\{E \} \backslash \{ R \}) \cup (\{ R \} \backslash \{ E \})$ (symmetric difference) is the resulting set of edges with bit flips left over after a round of noise following by corrections.
The symmetric difference is used since edges which are flipped by errors and also flipped back by the correction are not in a flipped state following the error correction round.
$\{\gamma\}$ is some path connecting the opposite edges of the surface code with no additional loops or disconnected components.
We categorize the edges $\{ M \}$ into two subsets: $\{ S \}$ and $\{ B \}$ for edges in the seam and bulk where errors occur with probabilities $p_\textrm{s}$ and $p_\textrm{b}$.
$\gamma_{S}$, $\gamma_{B}$ refer to the number of edges from these edge categories that intersect with the support of a given $\gamma$: $\gamma_S \equiv |\{\gamma\} \cap \{ S \}|$.
We will also refer to the numbers of seam edges contained in $\{\gamma\}$ and $\{E\}$, $\gamma_{SE} \equiv |\{ \gamma_S \} \cap \{E\}|$, and the number of seam edges contained in $\{\gamma\}$ and $\{R\}$, $\gamma_{SR} \equiv |\{ \gamma_S \} \cap \{R\}|$.

The syndrome measured given the set of errors $\{ E \}$ is $\partial \{ E \}$, which is the set of vertices adjacent to $\{ E \}$.
A given $\{E\}$ occurs with a probability
\begin{multline}
    \textrm{Prob}(\{ E \} ) = \prod_{i \in \{ E \} } p_i \prod_{i \in \{ M \} \backslash \{ E \} } (1 - p_i) \\
    \propto \prod_{i \in \{ E \} } \frac{p_i}{1-p_i}
    \label{chain_probs}
\end{multline}

MWPM then decodes $\partial \{ E \}$ and returns $\{R\}$, the set of edges the set of edges most likely to have resulted in the measured $\partial \{ E \}$, by maximizing  Eq.~\ref{chain_probs}, subject to the constraint $\partial \{ R \} = \partial \{ E \}$.
MWPM is an efficient classical algorithm known to perform nearly optimally at the task of making $\{E + R\}$ homologically trivial \cite{Higgott2021}. This is done by assigning weights $\log \big(\frac{1-p_i}{p_i} \big)$ to each edge in $\{ M \}$ and then finding the $\{R\}$ with the minimum weight such that $\partial \{ R \} = \partial \{ E \}$:
\begin{align}
    \textrm{wt} (\{ R \}) \equiv \sum_{i \in \{ R \} } \log \Big( \frac{1 - p_i}{p_i} \Big)
    \label{weight_def}
\end{align}

\subsection{Probability to fail via a particular $\{ \gamma \}$}

Our first goal is to upper bound the probability that a round of errors followed by corrective flips from MWPM will create a set of bit flips containing a particular $\{\gamma\}$ with $\gamma_{S}$ seam and $\gamma_{B}$ bulk edges and thus result in logical failure.
This amounts to bounding the probability that physical errors arise during a round of error correction such that $\{E + R\} \supseteq \{\gamma\}$ (i.e.~that the error chain $\{\gamma\}$ is contained within the resulting set of bit flips $\{E + R\}$). 

By definition, MWPM ensures $\textrm{wt}(\{ R \}) \leq \textrm{wt}(\{ E \})$.
For $\{E + R\} \supseteq \{\gamma \}$, since $\{\gamma\}$ is closed ($\partial \{\gamma\} = \emptyset$) it is then the case that $\{\gamma\} \cap \{R\}$ and $\{\gamma\} \cap \{E\}$ must share a boundary within $\{\gamma\}$.
Then $\textrm{wt}(\{\gamma\} \cap \{R\})  \leq \textrm{wt}(\{\gamma\} \cap \{E\})$, since if this were not the case, replacing the edges $\{\gamma\} \cap \{R\}$ by the edges of $\{\gamma\} \cap \{E\}$ would further minimize the weight of $\{R\}$, impossible by definition because MWPM already minimizes $\textrm{wt}(\{R\})$.

It is then the case that $\{E + R\} \supseteq \{\gamma \}$ implies the two conditions:
\begin{align}
    &\textrm{wt}( \{\gamma \}) = \textrm{wt}(\{E \} \cap \{\gamma \}) + \textrm{wt}(\{R\} \cap \{\gamma \}) \\
    &\textrm{wt}(\{R \} \cap \{\gamma \}) \leq \textrm{wt}(\{E \} \cap \{\gamma \})
    \label{mwpm_conditions}
\end{align}
and combining these gives that $ \textrm{wt}(\{E \} \cap \{\gamma \}) \geq \textrm{wt}(\{\gamma \})/2$.
Applying the definition of the weight Eq.~\ref{weight_def}, we then have the overall implication:
\begin{align}
    &\{E + R\} \supseteq \{\gamma \} \implies \label{weights_ineq} \\
    \Big(\frac{p_\textrm{s}}{1-p_\textrm{s}}\Big)^{\gamma_{SE}} & \Big(\frac{p_\textrm{b}}{1-p_\textrm{b}}\Big)^{\gamma_{BE}} \leq \Big(\frac{p_\textrm{s}}{1-p_\textrm{s}}\Big)^{\gamma_S/2} \Big(\frac{p_\textrm{b}}{1-p_\textrm{b}}\Big)^{\gamma_B/2} \nonumber
\end{align}

We will now bound $\textrm{Prob}(\gamma_{SE}, \gamma_{BE})$, the probability of generating an $\{E\}$ with exactly $\gamma_{SE}, \gamma_{BE}$ bit flips overlapping with $\{\gamma \}$.
This $\textrm{Prob}(\gamma_{SE}, \gamma_{BE})$ is the number of ways to choose $\gamma_{SE}$ from $\gamma_{S}$ and $\gamma_{BE}$ from $\gamma_{B}$, which we can bound since ${\gamma_S\choose \gamma_{SE} } \times { \gamma_B \choose \gamma_{BE} } \leq 2^{\gamma_S} \times 2^{\gamma_B}$, times the probability $p_\textrm{s}^{\gamma_{SE}} (1-p_\textrm{s})^{\gamma_S- \gamma_{SE}} p_\textrm{b}^{\gamma_{BE}} (1-p_\textrm{b})^{\gamma_B- \gamma_{BE}}$ to actually flip bits at each of those particular choices:
\begin{align}\label{prob_gamma}
\begin{split}
    &\textrm{Prob}(\gamma_{SE}, \gamma_{BE})\\ &\leq
     2^{\gamma_S+ \gamma_B} p_\textrm{s}^{\gamma_{SE}} (1-p_\textrm{s})^{\gamma_S- \gamma_{SE}} p_\textrm{b}^{\gamma_{BE}} (1-p_\textrm{b})^{\gamma_B- \gamma_{BE}}  \\
    &= 2^{\gamma_S+ \gamma_B} \Big(\frac{p_\textrm{s}}{1-p_\textrm{s}} \Big)^{\gamma_{SE}} (1-p_\textrm{s})^{\gamma_S} \Big( \frac{p_\textrm{b}}{1-p_\textrm{b}} \Big)^{\gamma_{BE}} (1-p_\textrm{b})^{\gamma_B}
\end{split}
\end{align}

Now we bound $\textrm{Prob}( \{E + R\} \supseteq \{\gamma \}\})$, the probability of an $\{ E \}$ occurring such that $\{E + R\} \supseteq \{\gamma \}$, allowing us to additionally impose the condition from Eq.~\ref{weights_ineq}, which we substitute into Eq.~\ref{prob_gamma} to obtain:
\begin{align}
    \textrm{Prob}(\{E + R\} & \supseteq \{\gamma \}) \leq  \nonumber \\
    2^{\gamma_S+ \gamma_B} \Big( \frac{p_\textrm{s}}{1-p_\textrm{s}} \Big)^{\gamma_S/2} & (1-p_\textrm{s})^{\gamma_S} \Big( \frac{p_\textrm{b}}{1-p_\textrm{b}} \Big)^{\gamma_B/2} (1-p_\textrm{b})^{\gamma_B} \nonumber \\
    & \leq 2^{\gamma_S+ \gamma_B} p_\textrm{s}^{\gamma_S/2} p_\textrm{b}^{\gamma_B/2} \label{gamma_bound}
\end{align}
where we have now successfully removed any explicit reference to the error set $\{ E \}$ and have bounded the failure probability only in terms of quantities $\gamma_S, \gamma_B$ dependent on the chosen $\{ \gamma \}$.

Since there can only be a logical failure if the round of noise followed by MWPM correction generates a set of bit flips containing some $\{ \gamma \}$ connecting opposite sides of the surface code, we can next bound the total logical failure probability $P_\textrm{fail}$ by summing the failure probability Eq.~\ref{gamma_bound} over all such $\{ \gamma \}$ connecting the two sides of the surface code:
\begin{align}\label{eqn:bound_gamma}
\begin{split}
    P_\textrm{fail} \leq \sum_{\{ \gamma \}} \textrm{Prob}( \{E + R\} \supseteq \{\gamma \}\}) \\
    \leq \sum_{ \{ \gamma \}} 2^{\gamma_S+\gamma_B} p_\textrm{s}^{\gamma_S/2} p_\textrm{b}^{\gamma_B/2}
\end{split}
\end{align}

\subsection{Counting nontrivial error chains}
The upper bound on $P_\mathrm{fail}$ given in \cref{eqn:bound_gamma} still depends on all possible $\{\gamma\}$.
In order to obtain an expression that we can evaluate more easily, we will instead sum over the larger set of self-avoiding walks (SAWs) of length $\ell \geq L$.
As all $\{\gamma\}$ are such SAWs, we can then obtain a still weaker upper bound on $P_\mathrm{fail}$.
To assign the correct probabilities, we will first discuss a parametrization of the SAWs that accounts for the number of bulk and seam edges contained in them.

Generally, we denote by $n_{\textrm{SAW}}$ the number of SAWs of some fixed length $\ell$.
To understand how $n_{\textrm{SAW}}$ grows with $\ell$, first consider growing a walk on a square lattice by appending edges in some dimension $D$.
Larger $D$ means more directions to choose every time you append an edge, so results in faster exponential growth of the number of walks.
For walks of length $\ell$, one bound is that:
\begin{equation}
    n_{\textrm{SAW}}(\ell) \leq (2D -1)^\ell
\end{equation}
as each new edge can be appended in any available direction other than back onto the walk itself.
The dimension then directly affects the threshold, as the faster that $n_{\textrm{SAW}}(\ell)$ blows up, the smaller the critical value of the error probabilities $p$ must be to control the growth of error chains and suppress the magnitude of $P_\textrm{fail}$ as the system size scales (as seen in Eq.~\ref{pfail_1}).

In our case, we need to count the number of walks as a function of $\gamma_{S}$ and the number of ways to insert bulk excursion segments hopping off and on the seam.
For walks with $\gamma_{S}$ seam edges we have ${\gamma_S \choose C}$ ways to choose $C$ locations (indexed by $k$) along the length of the walk to insert ``excursions", each of length $\ell_k$, where the walk jumps off the seam and into the bulk before rejoining the seam.
The excursion segments $\ell_k$, indexed from $k=1$ to $C$, together with bulk segments of length $\ell_0$ and $\ell_{C+1}$ at the beginning and end of the walk (if the walk starts or ends in the bulk instead of on the seam) form a set $\{ \ell_k \}_{k=0}^{C+1}$ which we will denote as just $\{ \ell_k \}$.
Once the locations of the corners are fixed along the length of the walk, they partition the walk into segments moving either in the bulk (dimension $D_\textrm{b}$) or seam (dimension $D_\textrm{s}$).
We count separately the two ``corner" bulk edges which project orthogonally from the seam at the beginning and end of each excursion, so that the total number of bulk edges resulting from the $C$ excursions is then $\gamma_B \equiv 2C + \sum_{k=0}^{C+1}\ell_k$.

Now for a fixed partitioning of the walk into seam and bulk segments with $\gamma_{S}$, $C$, $\{\ell_k \}$, we can follow the length of the walk and count the number of directions available in which to append edges when extending the walk in the seam or bulk as well as when encountering corners. When appending edges confined to the seam or during bulk excursions, we have $\mu_\textrm{s} \equiv 2D_\textrm{s}-1$ or $\mu_\textrm{b} \equiv 2D_\textrm{b}-1$ options. Each of the $C$ times the chain leaves the seam, there are $2(D_\textrm{b}-D_\textrm{s})$ corner edges to choose, and each of the $C$ times the chain moves back onto the seam from a corner, there are $2D_\textrm{s}$ options to choose the first seam edge of each seam segment (replacing the $\mu_\textrm{s}$ for those particular edges), so each excursion yields $2(D_\textrm{b}-D_\textrm{s}) \times (2D_\textrm{s}/\mu_\textrm{s})$ options, giving the overall bound when $C \geq 1$ and both seam and bulk edges are present:
\begin{align}\label{nsaw_C_notzero}
    n_\textrm{SAW}(\gamma_S,\{\ell_k \},C \geq 1) \leq {\gamma_S \choose C} a^{C} \mu_\textrm{s}^{\gamma_S} \mu_\textrm{b}^{\sum_k\ell_k}
\end{align}
where $a \equiv \mu_\textrm{c} / \mu_\textrm{s}$ and $\mu_\textrm{c} \equiv 4 D_\textrm{s} (D_\textrm{b}-D_\textrm{s})$ counts the number of ways to choose the two corner edges at the beginning and end of each excursion.
For $D_\textrm{s} = 2$, $D_\textrm{b} = 3$, we have $\mu_\textrm{s} = 3$, $\mu_\textrm{b} = 5$ and $a =\frac{8}{3}$. Only slightly smaller values than this simple counting argument for $\mu_\textrm{s}, \mu_\textrm{b}$ are known \cite{Dennis2002,slade_1996}.

If $C=0$, there is no SAW with both seam and bulk edges and the walk is only in either bulk or seam so $S=0$ or $B=0$:
\begin{align}\label{nsaw_C_zero}
    n_\textrm{SAW}(\gamma_S,|\{\ell_k \}|=0,C=0) \leq \mu_\textrm{s}^{\gamma_S} \nonumber \\
    n_\textrm{SAW}(\gamma_S=0,\{\ell_k \},C=0) \leq \mu_\textrm{b}^{\gamma_B}
\end{align}

\subsection{Bounding the logical failure probability}
Let $\textrm{Prob}_\textrm{SAW}(\gamma_S,\{\ell_k \},C)$ be the total probability to generate an error set $\{E\}$ such that any $\{\gamma\}$ with numbers of edges $\gamma_{S}$, $\{\ell_k \}$, and $C$ is contained in $\{E + R\}$.
As all $\{\gamma\}$ with the same $\gamma_S,\{\ell_k \},C$ occur with the same probability bound given by Eq.~\ref{gamma_bound}, we can express $\textrm{Prob}_\textrm{SAW}(\gamma_S,\{\ell_k \},C)$ as the probability from Eq.~\ref{gamma_bound} times the number of such walks $n_\textrm{SAW}(\gamma_S,\{\ell_k \},C)$.
As there are $\textrm{poly}(L)$ possible edges to start the walk in the matching graph, we have
\begin{multline}
    \textrm{Prob}_\textrm{SAW}(\gamma_S,\{\ell_k \},C)/\textrm{poly}(L) \leq \\
    n_\textrm{SAW}(\gamma_S,\{\ell_k \},C)
    \times 2^{\gamma_S+ [2C + \sum_k\ell_k]} p_\textrm{s}^{\gamma_S/2} p_\textrm{b}^{[2C + \sum_k\ell_k]/2}
\end{multline}

Then we can bound the logical failure probability $P_\textrm{fail}$, since forming homologically non-trivial loops requires error chain walks with at least $L$ edges stretching in the space or time direction along the seam so that the number of combined seam and bulk excursion edges is sufficiently large: $\gamma_S+\sum_{k=0}^{C+1}\ell_k \geq L$.
Corner edges themselves do not contribute to generating walks along a direction necessary for failure.
Summing over all the walks with values of $\gamma_S,\{\ell_k \},C$ which could contribute to logical failure gives:
\begin{multline}
    P_\textrm{fail}/\textrm{poly}(L) \\
    \leq \sum_{\gamma_S\geq 0} \sum_{C = 0}^{\gamma_S} \sum_{\substack{\{\ell_k  \geq 1 \}: \\ \gamma_S+\sum_k\ell_k \geq L}} \textrm{Prob}_\textrm{SAW}(\gamma_S,C,\{\ell_k \})\\
    \leq \sum_{\gamma_S\geq 0} \sum_{C = 0}^{\gamma_S} \sum_{\substack{\{\ell_k  \geq 1 \}: \\ \gamma_S+\sum_k\ell_k \geq L}} n_\textrm{SAW}(\gamma_S,C,\{\ell_k \}) \\
    \times (4p_\textrm{s})^{\gamma_S/2} (4 p_\textrm{b})^{\frac{1}{2} \sum_k\ell_k} (4 p_\textrm{b})^{C}
    \label{pfail_1}
\end{multline}

\begin{figure*}
\includegraphics[width=18cm]{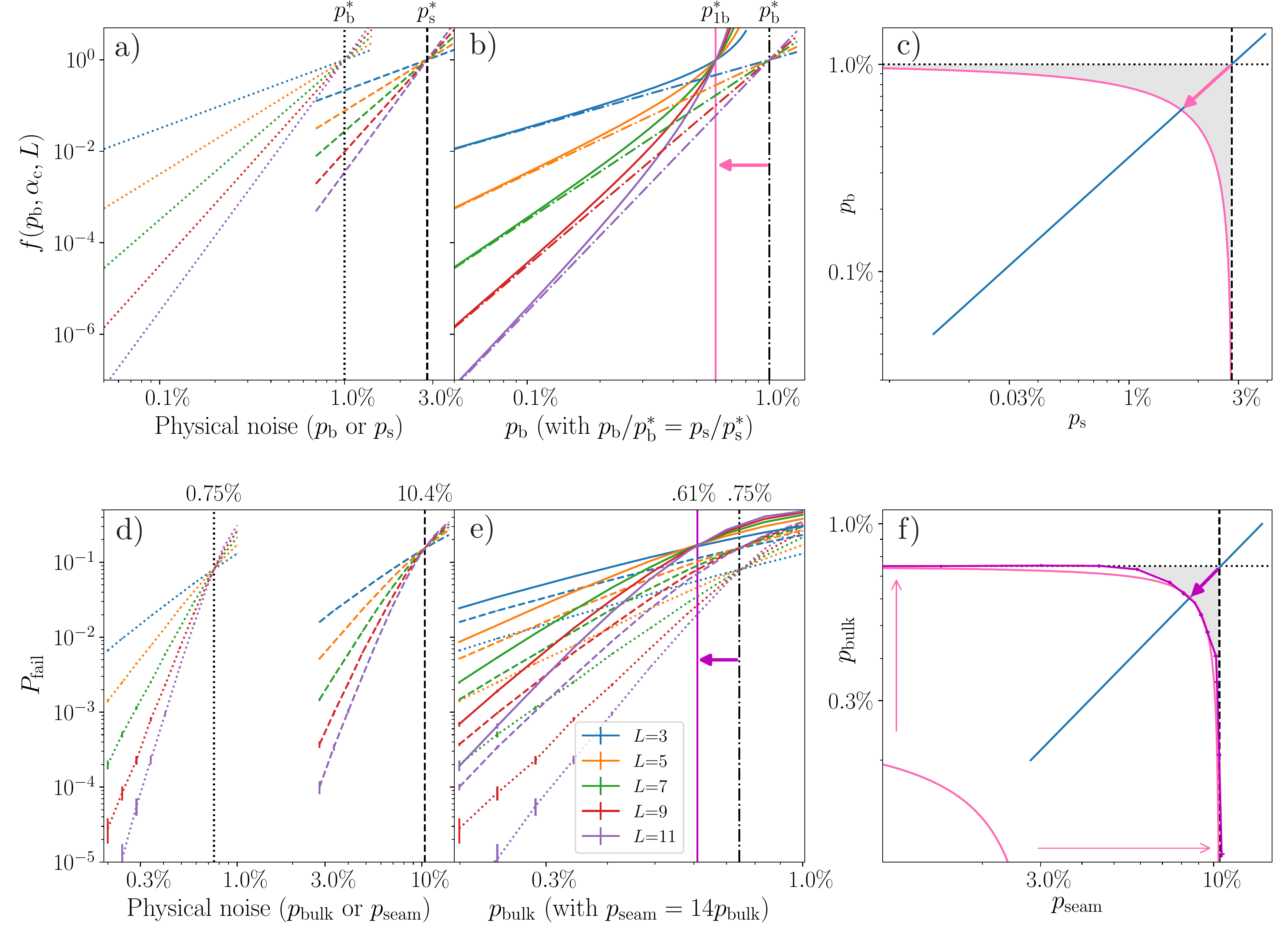}
\caption{
Top row: analytical logical failure bounds.
a) Bounds for $D_\textrm{b} = 3$ (dotted) and $D_\textrm{s} = 2$ (dashed), with threshold bounds $p_\textrm{b}^*$ and $p_\textrm{s}^*$ respectively.
b) Analytical bounds (Eq.~\ref{Pfail_equal}) fixing $p_\textrm{b}/p_\textrm{b}^* = p_\textrm{s}/p_\textrm{s}^*$ (solid), in which case the seam and bulk curves from a) now overlap (dot-dashed) and we plot them vs $p_\textrm{b}$. Seam-bulk interactions reduce the threshold bound slightly to $p_{1b}^*$ as indicated by the pink arrow.
The logical failure rate converges to the values for no seam-bulk interactions once a few times below threshold as excursions into the bulk become ``frozen out."
c) Plotting the threshold bound Eq.~\ref{thresh_formula} (pink) in the space of possible choices for $p_\textrm{s}, p_\textrm{b}$. In the absence of seam-bulk interactions, all points to the left of the dashed line (so $p_\textrm{s} < p_\textrm{s}^*)$ and below the dotted line (so $p_\textrm{b} < p_\textrm{b}^*$) would be below the threshold bound. Seam-bulk interactions make the threshold bound sag to the pink curve, where the gray region is no longer below the threshold bound.
The blue line shows the cut corresponding to the horizontal axis from b).
Bottom row: numerical simulations.
d) Same as a) but exact numerical simulation.
e) Same as b) but exact numerical simulation with the choice $p_\textrm{seam} = 14 p_\textrm{bulk}$, approximately $p_\textrm{bulk}/p_\textrm{bulk}^* = p_\textrm{seam}/p_\textrm{seam}^*$.
Curves including seam-bulk interaction (solid) similarly converge toward the seam-only curves (dashed) as $p_\textrm{bulk}/p_\textrm{bulk}^*$ becomes small.
f) Numerically extracted threshold plotted in terms of $p_\textrm{seam}, p_\textrm{bulk}$ (purple). The pink curve is the bound replotted from c).
Light pink shows the bound with numerically extracted thresholds substituted in along with an effective value of $\alpha_\textrm{c} \rightarrow 1.4$, the minimal value which still bounds all the numerical datapoints.
}
\label{threshold_shifts}
\end{figure*}

Plugging in our expressions Eq.~\ref{nsaw_C_notzero} and \ref{nsaw_C_zero} for $n_\textrm{SAW}$ into Eq.~\ref{pfail_1}, we get terms corresponding to purely bulk chains ($S=0$) and purely seam chains ($B=0$), with an additional sum over walks across the seam and bulk when there is at least one excursion $C$:
\begin{multline}
    P_\textrm{fail}/\textrm{poly}(L) \leq \sum_{\gamma_S\geq L} \Big(\frac{p_\textrm{s}}{p_\textrm{s}^*}\Big)^{\frac{\gamma_S}{2}} + \sum_{\gamma_B\geq L} \Big(\frac{p_\textrm{b}}{p_\textrm{b}^*}\Big)^{\frac{\gamma_B}{2}} + \\
    \sum_{\gamma_S\geq 1} \Big(\frac{p_\textrm{s}}{p_\textrm{s}^*}\Big)^{\frac{\gamma_S}{2}}
    \sum_{C = 1}^{\gamma_S} {\gamma_S\choose C} (4a p_\textrm{b})^{C} \sum_{\substack{\{\ell_k \geq 1 \}: \\ \gamma_S+\sum_k\ell_k \geq L}} \Big(\frac{p_\textrm{b}}{p_\textrm{b}^*}\Big)^{\frac{1}{2}\sum_k\ell_k}
    \label{pfail_intermediate}
\end{multline}
where $p_\textrm{s}^* \equiv \frac{1}{4 \mu_\textrm{s}^2}$, $p_\textrm{b}^* \equiv \frac{1}{4 \mu_\textrm{b}^2}$ correspond to bounds on the thresholds for the seam and bulk when treated independently as in Fig.~\ref{threshold_shifts}a. For the case of interest where $D_\textrm{s} = 2, D_\textrm{b} = 3$ we have values $p_\textrm{s}^* \equiv \frac{1}{4 \mu_\textrm{s}^2} = \frac{1}{4 \times 3^2} = 0.028$, $p_\textrm{b}^* \equiv \frac{1}{4 \mu_\textrm{b}^2} = \frac{1}{4 \times 5^2} = 0.010$ (see Fig.~\ref{threshold_shifts}a).

Now we can use the constraint $\sum\ell_k \geq L-\gamma_S$ to realize that these geometric sums cannot all start at $l_k=1$ when $\gamma_S < L$. We denote that they start at some $d_k$ such that $\sum_k d_k \equiv L-\gamma_S$ for $L -\gamma_S> 0$ and $\sum_k d_k \equiv 0$ for $L -\gamma_S\ngeq 0$ (which we denote as $\sum_k d_k \equiv L -\gamma_S|_{>0}$) and we then substitute $l_k =\ell_k' + d_k$ so that $l_k'$ now represents the length of the ``slack" in the chain at each excursion:
\begin{multline}
    \sum_{\substack{\{\ell_k \geq 1 \}: \\ \gamma_S+\sum_k\ell_k \geq L}} \Big(\frac{p_\textrm{b}}{p_\textrm{b}^*}\Big)^{\frac{1}{2}\sum_k\ell_k}  = 
    \sum_{\substack{\{\ell_k' \geq 0 \}}}  \Big(\frac{p_\textrm{b}}{p_\textrm{b}^*}\Big)^{\frac{1}{2}\sum_k d_k +\ell_k'} \\
    = \Big(\frac{p_\textrm{b}}{p_\textrm{b}^*}\Big)^{\frac{L-\gamma_S}{2} \big|_{>0}} \sum_{\substack{\{\ell_k' \geq 0 \}}} \Big(\frac{p_\textrm{b}}{p_\textrm{b}^*}\Big)^{\frac{1}{2}\sum_k\ell_k'}
\end{multline}
where the same substitution transforms the constraint $\gamma_S+\sum_k\ell_k \geq L \rightarrow \gamma_S+\sum_k\ell_k' + \sum_k d_k \geq L \implies \sum_k\ell_k' \geq 0$ (using $\sum_k d_k = L -\gamma_S|_{>0}$), which is then trivial and we drop. For simplicity we now relabel the $l_k'$ as just $l_k$ again.

The true upper limit on the sum would be determined when the volume of the bulk was filled by the excursions, so up to the constraint $\sum_k\ell_k \leq \textrm{poly}(L)$. Then it is also the case that we can maintain a bound by extending the sum up to the constraint $l_k \leq \textrm{poly}(L)$ for each $\ell_k$.
Recall how we defined the notation $\{\ell_k\} = \{\ell_k\}_{k=0}^{C+1}$. We can then separate out the sums over the chain segment lengths of $\ell_0$ and $\ell_{C+1}$, as for $\frac{p_\textrm{b}}{p_\textrm{b}^*} < 1$ it is then the case that $\sum_{\ell_0 \geq 0}^{\textrm{poly}(L)}\Big(\frac{p_\textrm{b}}{p_\textrm{b}^*}\Big)^{\ell_0/2} \leq \textrm{poly}(L)$ and $\sum_{\ell_{C+1} \geq 0}^{\textrm{poly}(L)}\Big(\frac{p_\textrm{b}}{p_\textrm{b}^*}\Big)^{\ell_{C+1}/2} \leq \textrm{poly}(L)$ since all terms in the sum are less than $1$, and this can be absorbed into $P_\textrm{fail}/\textrm{poly}(L)$. 
Now, the remaining set of sections $\{ \ell_k \}$ runs from $k=1$ to $C$.
Intuitively, since the bulk segments of length $\ell_0$ and $\ell_{C+1}$ at the beginning and end of the error chain each only occur once, they do not affect the threshold bound, whereas the number $C$ of possible seam excursions in the middle of the error chain can scale with the system size, and so will affect the threshold.

To deal with the excursions, we can simplify and further maintain a bound by extending the upper limit to allow each $l_k$ to run up to $\infty$ independently:
\begin{align}
    \sum_{\{\ell_k \geq 0\}_{k=1}^{C}}^{\textrm{poly}(L)} \Big(\frac{p_\textrm{b}}{p_\textrm{b}^*}\Big)^{\frac{1}{2}\sum_k\ell_k} \leq \sum_{\{\ell_k \geq 0\}_{k=1}^{C}}^\infty \Big(\frac{p_\textrm{b}}{p_\textrm{b}^*}\Big)^{\frac{1}{2}\sum_k\ell_k}
\end{align}

Next we can factor out the sum across the set of values $\{\ell_k \}_{k=1}^{C}$ into a product and use the geometric sum $\sum_{n=0}^\infty r^n = \frac{1}{1-r}$ assuming $\frac{p_\textrm{b}}{p_\textrm{b}^*} < 1$:
\begin{multline}
    \sum_{\{\ell_k \geq 0\}_{k=1}^{C}}^\infty \Big(\frac{p_\textrm{b}}{p_\textrm{b}^*}\Big)^{\frac{1}{2}\sum_k\ell_k} = \prod_{k=1}^C \Big[\sum_{\ell_k = 0}^\infty  \Big(\frac{p_\textrm{b}}{p_\textrm{b}^*}\Big)^{\frac{1}{2}\ell_k} \Big] = \\
    \Bigg[ \sum_{\ell = 0}^\infty  \Big(\frac{p_\textrm{b}}{p_\textrm{b}^*}\Big)^{\frac{1}{2} \ell} \Bigg]^C = \Bigg[ \frac{1}{1-\sqrt{p_\textrm{b}/p_\textrm{b}^*}} \Bigg]^C
\end{multline}

Substituting this into \ref{pfail_intermediate} we arrive at:
\begin{multline}
    P_\textrm{fail}/\textrm{poly}(L) \leq \sum_{\gamma_S\geq L} \Big(\frac{p_\textrm{s}}{p_\textrm{s}^*}\Big)^{\frac{\gamma_S}{2}} + \sum_{\gamma_B\geq L} \Big(\frac{p_\textrm{b}}{p_\textrm{b}^*}\Big)^{\frac{\gamma_B}{2}} \\
    + \sum_{\gamma_S\geq 1} \Big(\frac{p_\textrm{s}}{p_\textrm{s}^*}\Big)^{\gamma_S/2} \Big(\frac{p_\textrm{b}}{p_\textrm{b}^*}\Big)^{\frac{L-\gamma_S}{2} \big|_{>0}}\\
    \times \sum_{C = 1}^{\gamma_S} {\gamma_S\choose C} (4a p_\textrm{b})^{C} \Bigg[ \frac{1}{1-\sqrt{p_\textrm{b}/p_\textrm{b}^*}} \Bigg]^C
\end{multline}

Now we can use the identity $\sum_{i=0}^n {n \choose i}x^i = (1+x)^n$:
\begin{multline}
    P_\textrm{fail}/\textrm{poly}(L) \leq \sum_{\gamma_S\geq L} \Big(\frac{p_\textrm{s}}{p_\textrm{s}^*}\Big)^{\frac{\gamma_S}{2}} + \sum_{\gamma_B\geq L} \Big(\frac{p_\textrm{b}}{p_\textrm{b}^*}\Big)^{\frac{\gamma_B}{2}} + \\
    \sum_{\substack{\gamma_S\geq 1: \\ C \neq 0}} \Big(\frac{p_\textrm{s}}{p_\textrm{s}^*}\Big)^{\frac{\gamma_S}{2}} \Big(\frac{p_\textrm{b}}{p_\textrm{b}^*}\Big)^{\frac{L-\gamma_S}{2} \big|_{>0}} \Bigg[1 + \alpha_\textrm{c} p_\textrm{b} \frac{\sqrt{p_\textrm{s}^*}}{1-\sqrt{p_\textrm{b}/p_\textrm{b}^*}} \Bigg]^{\gamma_S}
\end{multline}
where $4a = \alpha_c \sqrt{p_s^*}$, where we now think of $\alpha_c$ as effectively counting the number of ways to leave and return to the seam on a particular excursion.

Now, if we have no excursions $C = 0$:
\begin{align}
    P_\textrm{fail}(C = 0)/\textrm{poly}(L) \leq
    \sum_{\gamma_S\geq L} \Big(\frac{p_\textrm{s}}{p_\textrm{s}^*} \Big)^{\frac{\gamma_S}{2}} + \sum_{\gamma_B\geq L} \Big(\frac{p_\textrm{b}}{p_\textrm{b}^*}\Big)^{\frac{\gamma_B}{2}}
\end{align}
Now for $\frac{p_\textrm{s}}{p_\textrm{s}^*} < 1, \frac{p_\textrm{b}}{p_\textrm{b}^*} < 1$, the largest terms in the sums of $\gamma_S,\gamma_B$ will be the ones where $\gamma_S=L$ and $\gamma_B=L$, and as the sums can only have at maximum a number of terms equal to lattice size, some $\textrm{poly}(L)$, we have:
\begin{align}
    P_\textrm{fail}(C = 0)/\textrm{poly}(L) \leq
    \Big(\frac{p_\textrm{s}}{p_\textrm{s}^*} \Big)^{\frac{L}{2}} + \Big(\frac{p_\textrm{b}}{p_\textrm{b}^*}\Big)^{\frac{L}{2}}
\end{align}

A similar argument applies to the cross terms:
\begin{multline}
    P_\textrm{fail}/\textrm{poly}(L) \leq
    \Big(\frac{p_\textrm{s}}{p_\textrm{s}^*} \Big)^{\frac{L}{2}} + \Big(\frac{p_\textrm{b}}{p_\textrm{b}^*}\Big)^{\frac{L}{2}} + \\
    \sum_{\substack{ \gamma_S \geq 1: \\ C \neq 0}}^{L} \Bigg[ \frac{p_\textrm{s}}{p_\textrm{s}^*}\Big(1 + \alpha_\textrm{c} p_\textrm{b} \frac{\sqrt{p_\textrm{s}^*}}{1-\sqrt{p_\textrm{b}/p_\textrm{b}^*}}\Big)^2 \Bigg]^{\frac{\gamma_S}{2}} \Big(\frac{p_\textrm{b}}{p_\textrm{b}^*}\Big)^{\frac{L-\gamma_S}{2}}
    \label{Pfail_cross}
\end{multline}
which are all suppressed as $L \rightarrow \infty$ provided that
\begin{align}
    \frac{p_\textrm{s}}{p_\textrm{s}^*} \Bigg[1 + \alpha_\textrm{c} p_\textrm{b} \frac{\sqrt{p_\textrm{s}^*}}{1-\sqrt{p_\textrm{b}/p_\textrm{b}^*}} \Bigg]^2 < 1 
    \label{thresh_formula} \\
    \frac{p_\textrm{b}}{p_\textrm{b}^*} < 1
\end{align}

Re-expressing Eqn. \ref{thresh_formula},
we can see that it is equivalent to a small downward ``sag" of the threshold bound:
\begin{equation}
    p_\textrm{s}^* \rightarrow p_\textrm{1s}^* \equiv p_\textrm{s}^* \bigg/ \Bigg[1 + \alpha_\textrm{c} p_\textrm{b} \frac{\sqrt{p_\textrm{s}^*}}{1-\sqrt{p_\textrm{b}/p_\textrm{b}^*}} \Bigg]^{2}
    \label{thresh_formula_si}
\end{equation}
demonstrating a tradeoff between different amounts of ``sag" in the seam and bulk thresholds when both seam and bulk errors are occurring (see Fig.~\ref{threshold_shifts}c).

Specializing to the case $\frac{p_\textrm{s}}{p_\textrm{s}^*} = \frac{p_\textrm{b}}{p_\textrm{b}^*}$ we can further bound and simplify, which we plot in Fig.~\ref{threshold_shifts}b as well as Fig.~3b of the main text and compare to numerical simulations:
\begin{multline}
    P_\textrm{fail}/\textrm{poly}(L) \leq \\
    \Bigg[ \frac{p_\textrm{b}}{p_\textrm{b}^*}\Big(1 + \alpha_\textrm{c} p_\textrm{b} \frac{\sqrt{p_\textrm{s}^*}}{1-\sqrt{p_\textrm{b}/p_\textrm{b}^*}}\Big)^2 \Bigg]^{\frac{L}{2}} \equiv f(p_\textrm{b}, \alpha_\textrm{c}, L)
    \label{Pfail_equal}
\end{multline}

Eq.~\ref{Pfail_equal} suggests not only what combinations of seam and bulk errors achieve a threshold but also, importantly for determining the scaling efficiency, that the subthreshold behavior quickly approaches that of independent bulk and seam when the bulk itself is subthreshold.
From the numbers above, we can see that with $p_\textrm{b}$ just a few times below $p_\textrm{b}^*$, excursions from the seam into the bulk are strongly suppressed, so that the subthreshold scaling of the seam terms in Eq.~\ref{Pfail_cross}  is nearly the same as without any bulk errors.
Indeed, we see behavior consistent with this observation in the numerical results in Fig.~\ref{threshold_shifts}, with the curves for combined seam and bulk errors approaching those for purely seam errors once a few times below threshold.

\subsection{Extending the model to 2 seams}
In this section we generalize our approach to understand the case of having two seams embedded in the same bulk, where error chains can now hop between seams in addition to hopping off and on the same seam.

We can interpret our results from the previous section in the following way. Increasing $\gamma_{S}$ by 1 to append an additional seam edge leads to an additional factor of:
\begin{align}
    \sqrt{\frac{p_\textrm{s}}{p_\textrm{s}^*} } \Bigg[1 + \alpha_\textrm{c} p_\textrm{b} \frac{\sqrt{p_\textrm{s}^*}}{1-\sqrt{p_\textrm{b}/p_\textrm{b}^*}} \Bigg]
\end{align}
where this factor must be smaller than unity to be below threshold.
To interpret our results, we can rewrite the factor from above:

\begin{align}
    =  \mu_\textrm{s} \sqrt{4 p_\textrm{s}}  +  \mu_\textrm{c} \sqrt{4 p_\textrm{s}} ( \sqrt{4 p_\textrm{b}})^2 \frac{1}{1-\sqrt{p_\textrm{b}/p_\textrm{b}^*}} \nonumber\\
    = \mu_\textrm{s} \sqrt{4 p_\textrm{s}}  +  \mu_\textrm{c} \sqrt{4 p_\textrm{s}} ( \sqrt{4 p_\textrm{b}})^2 \sum_{\ell=0}^\infty \Big(\frac{p_\textrm{b}}{p_\textrm{b}^*}\Big)^{\ell/2} \nonumber \\ 
    = \mu_\textrm{s} \sqrt{4 p_\textrm{s}}  +  \mu_\textrm{c} \sqrt{4 p_\textrm{s}} (\sqrt{4 p_\textrm{b}})^2 \sum_{\ell=0}^\infty \Big( \mu_\textrm{b} \sqrt{4 p_\textrm{b}} \Big)^{\ell}
\end{align}

We interpret this factor as a sum across the different ways possible to append the next seam edge, weighted by the ``probabilities" associated with each edge (in fact the square roots of the ``probabilities" because of the argument from above where MWPM can fill in missing edges).
The next seam edge can be added either by remaining on the seam and locally appending another seam edge ($\sqrt{4 p_\textrm{s}}$ with  $\mu_\textrm{s}$ options), or by first jumping out into the bulk, appending $\ell$ bulk edges, and then reattaching back onto the seam ($\sqrt{4 p_\textrm{s}} (\sqrt{4 p_\textrm{b}})^2$ with $\mu_{c}$ ways to establish the beginning and end points of the seam and summed over all ways to have $k$ bulk edges in the middle of the excursion).

This kind of approach is helpful to understand not only how excursions from a single seam back onto itself behave, but also other situations, such as when you have multiple seams within a code.
For example, when there are two seams it would make sense to consider additional terms to the sum above to account for excursions which hop between seams in addition to hopping off/on the same seam.
In that case, we can qualitatively add a term corresponding to all the paths which leave one of the seams and rejoin the opposite seam:
\begin{align}
    (\mu_{2c} \sqrt{4 p_\textrm{s}} \sqrt{4 p_\textrm{b}}^2) \sum_{\ell=h-2}^\infty \Big( \mu_\textrm{b} \sqrt{4 p_\textrm{b}} \Big)^{\ell}
\end{align}
appending chains of $\ell$ bulk edges to hop between seams, where $h$ is the distance separating the seams (and the sum is from $h-2$ because there have to be at least that many bulk edges in addition to the corner edges to stretch between the seams), and $\mu_{2c}$ is the number of ways to leave one seam and rejoin the second ($\mu_{2c} = 4$ is a reasonable choice since these paths would be dominated by ones which leave in the direction toward the other seam, so one choice to leave and 4 to rejoin).
We then have the total sum:
\begin{align}
    \mu_\textrm{s} \sqrt{4 p_\textrm{s}}  +  (\mu_{c} \sqrt{4 p_\textrm{s}} \sqrt{4 p_\textrm{b}}^2) \sum_{\ell=0}^\infty \Big( \mu_\textrm{b} \sqrt{4 p_\textrm{b}} \Big)^{\ell} \nonumber \\
    + (\mu_{2c} \sqrt{4 p_\textrm{s}} \sqrt{4 p_\textrm{b}}^2) \sum_{\ell=h-2}^\infty \Big( \mu_\textrm{b} \sqrt{4 p_\textrm{b}} \Big)^{\ell}
\end{align}

As done in the previous section, we can collapse the threshold shift due to a single seam: $p_\textrm{s}^* \rightarrow p_{1s}^{*}$ and then have only the interseam hops separated out:
\begin{align}
    \sqrt{\frac{p_\textrm{s}}{p_\textrm{1s}^{*}}} + \mu_\textrm{2c} \sqrt{4 p_\textrm{s}} \sqrt{4 p_\textrm{b}}^2 \sum_{\ell=h-2}^\infty \Big( \mu_\textrm{b} \sqrt{4 p_\textrm{b}} \Big)^{\ell} \nonumber \\
    = \sqrt{\frac{p_\textrm{s}}{p_\textrm{1s}^{*}}} \Bigg[1 + \alpha_\textrm{2c} \sqrt{ p_\textrm{1s}^{*}} \sqrt{p_\textrm{b}}^2 \bigg(\frac{p_\textrm{b}}{p_\textrm{b}^*} \bigg)^{(h-2)/2} \frac{1}{1-\sqrt{p_\textrm{b}/p_\textrm{b}^*}} \Bigg] \nonumber
\end{align}

Like before we can move the whole term in brackets down against $p_{1s}^{*}$ to find the form of the expected additional threshold shift due to hops between seams:
\begin{align}
    p_\textrm{1s}^{*} \rightarrow p_\textrm{2s}^{*} \equiv
    p_\textrm{1s}^{*} \bigg/ \Bigg[1 + \alpha_\textrm{2c} p_\textrm{b} \frac{\sqrt{p_\textrm{1s}^{*}} }{1-\sqrt{p_\textrm{b}/p_\textrm{b}^*}} \bigg(\frac{p_\textrm{b}}{p_\textrm{b}^*} \bigg)^{(h-2)/2} \Bigg]^2
    \label{2_seam_sag}
\end{align}

Notice that the extra threshold shift here disappears for large $h$; long hops through the bulk between seams are then highly suppressed and become a negligible contributor to the ways to add the next seam edge.

Choosing for example $p_\textrm{b}/p_\textrm{b}^* \rightarrow \frac{1}{2}$, from main text Fig.~3 we can see that there is only a slight drop in the seam error threshold due to having the bulk surrounding a single seam. Setting $p_{1s}^*$ to this value, in main text Fig.~4 we plot Eq.~\ref{2_seam_sag} for an effective value of $\alpha_{2c} = 6.1$, which then closely matches the behavior of the numerically extracted seam error threshold for 2 seams.
The analytical formulas appear to capture the qualitative relationship between the distance between seams and how far below threshold the bulk is, and can easily be generalized to cases of having many seams and in higher dimensions, such as building a large surface code from smaller patches.

\bibliography{ramette_biblio}